\documentclass[twocolumn]{aastex62}

\submitjournal{ApJ}

\shorttitle{Dynamics of the Convective Turbulence}
\shortauthors{Ishikawa et al.}

\newcommand{\ilam}{I_{\lambda}}
\newcommand{\hati}{\hat{I}}
\newcommand{\icont}{I_{\mathrm{c}}}
\newcommand{\ptot}{P_{\mathrm{tot}}}

\newcommand{\FWfif}{\mathrm{FWHM}}

\newcommand{\Dopsev}{v_{0.7}}
\newcommand{\Dopfif}{v_{0.5}}

\newcommand{\Dopofi}{v_{0.05}}
\newcommand{\veldif}{\Delta v}

\begin{document}

\title{Study of the Dynamics of Convective Turbulence in the Solar Granulation\\
by Spectral Line Broadening and Asymmetry}

\correspondingauthor{Ryohtaroh T. Ishikawa}
\email{ryohtaroh.ishikawa@nao.ac.jp}

\author[0000-0002-4669-5376]{Ryohtaroh T. Ishikawa}
\affiliation{Department of Astronomical Science, School of Physical Sciences, The Gradiate University for Advanced Studies, SOKENDAI, 2-21-1 Osawa, Mitaka, Tokyo 181-8588, Japan}
\affiliation{National Astronomical Observatory of Japan, 2-21-1 Osawa, Mitaka, Tokyo 181-8588, Japan}

\author[0000-0002-5054-8782]{Yukio Katsukawa}
\affiliation{National Astronomical Observatory of Japan, 2-21-1 Osawa, Mitaka, Tokyo 181-8588, Japan}
\affiliation{Department of Astronomical Science, School of Physical Sciences, The Gradiate University for Advanced Studies, SOKENDAI, 2-21-1 Osawa, Mitaka, Tokyo 181-8588, Japan}

\author{Takayoshi Oba}
\affiliation{Institute of Space and Astronautical Science, Japan Aerospace Exploration Agency, 3-1-1 Yoshinodai, Chuo-ku, Sagamihara, Kanagawa 252–5210, Japan}

\author[0000-0002-5892-6047]{Motoki Nakata}
\affiliation{National Institute for Fusion Science, 322-6 Oroshi-cho, Toki, Gihu 509-5292, Japan}

\author{Kenichi Nagaoka}
\affiliation{National Institute for Fusion Science, 322-6 Oroshi-cho, Toki, Gihu 509-5292, Japan}
\affiliation{Nagoya University, Huro-cho, Chikusa-ku, Nagoya, Aichi 464-8601, Japan}

\author{Tatsuya Kobayashi}
\affiliation{National Institute for Fusion Science, 322-6 Oroshi-cho, Toki, Gihu 509-5292, Japan}



\

\begin{abstract}
In the quiet regions on the solar surface,
turbulent convective motions of granulation play an important role in creating small-scale magnetic structures,
as well as in energy injection into the upper atmosphere.
The turbulent nature of granulation can be studied using spectral line profiles, especially line broadening,
which contains information on the flow field smaller than the spatial resolution of an instrument.
Moreover, the Doppler velocity gradient along a line-of-sight (LOS) causes line broadening as well.
However, the quantitative relationship between velocity gradient and line broadening has not been understood well.
In this study, we perform bisector analyses using the spectral profiles
obtained using the Spectro-Polarimeter of the Hinode/Solar Optical Telescope to investigate
the relationship of line broadening and bisector velocities with the granulation flows.
The results indicate that line broadening has a positive correlation with the Doppler velocity gradients along the LOS.
We found excessive line broadening in fading granules, that cannot be explained only by the LOS velocity gradient,
although the velocity gradient is enhanced in the process of fading.
If this excessive line broadening is attributed to small-scale turbulent motions, the averaged turbulent velocity is obtained as 0.9 km/s.
\end{abstract}



\section{Introduction} \label{sec:intro}
Regions of low solar activity, known as quiet regions of the Sun, are dominated by bright cells termed granules,
which are surrounded by dark channels called intergranular lanes.
Granules are caused by gas convection:
hot ascending material makes a granule and cold descending material makes an intergranular lane.
The behavior of granules is very dynamic that they undergo a cycle of birth, death, fragmentation, and merger.
The physical properties and the temporal evolutions of granules have been reported in numerous studies
(\citealt{Hirzberger97, Roudier86, Roudier03, Oba17a, Oba17b}).
Some small granules fade out without any fragmentation or merger,
whereas large granules principally fragment into multiple smaller granules \citep{Hirzberger99, Muller01, Lemmerer17}.

Small-scale magnetic fields are ubiquitous in such quiet regions (recent review by \citealt{BellotRubio19}).
Such magnetic fields are the energy source of coronal heating and solar wind acceleration (see \citealt{Cranmer19} and references therein).
Several observations have found that the internetwork magnetic fields have no clear correlation with
the solar cycle \citep{Hagenaar03, Buehler13, Lites14},
which implies that internetwork fields are probably created by
\deleted{a }small-scale \replaced{dynamo}{dynamos} driven by surface convection
rather than the global dynamo mechanism.
Such small-scale dynamos have been found in recent numerical simulations \citep{Shelyag11, Rempel14}.
In addition, small-scale vortex motions in the photosphere \citep{Bonet08, VargasDominguez11} could be a driver of small-scale dynamos.
Thus, studying granular- and subgranular-scale flows in the photosphere is important.

\citet{Abramenko12} found that two categories of granules exist: regular-size and {\it mini}-size granules.
They suggested that the mini-granules behave in a turbulent manner compared with their regular-size counterparts.
\citet{vanKooten17} analyzed the motions of bright points in a numerically simulated photosphere
and concluded that mini-granules cause high-frequency motions
and regular-size granules cause low-frequency motions.
They further suggested that there might be more power observed \replaced{at}{in} small-scale and high-frequency phenomena.
However, obtaining the velocity fields at such a small scale is difficult with the current instruments.

One possible approach to derive the velocity fields smaller than the spatial resolution is to use spectral line broadening.
\citet{Nesis92} found an anti-correlation between the line width and the continuum intensity,
suggesting that shocks are excited by collisions of supersonic horizontal flows.
\citet{Rybak04} postulated that the increase in \replaced{line core}{a line-core} intensity associated with the transient increase in line width
indicates the heating induced by post-shock turbulence.
However, \citet{Solanki96} found that such anti-correlations can be created
without any shocks by simulating granulation and studying synthesized spectral signatures.

Furthermore, \added{a }Doppler velocity gradient along the line-of-sight (LOS) causes spectral line broadening as well.
\citet{Gadun97} performed 2-D hydrodynamic numerical simulations
and calculated the spectral line widths of synthesized spectral lines.
The simulated photosphere creates velocity gradients along the height direction, especially in the intergranular lanes. 
Moreover, the synthesized spectral lines exhibit a broad profile. 
\citet{Khomenko10} measured the spatial distribution of the spectral line widths
obtained with a filtergraph onboard a balloon-borne telescope (SUNRISE/IMaX; \citealt{Barthol11,Martinez11}).
They found that the spectral line width becomes larger in intergranular downflow lanes
and becomes very small at the boundary between a granule and an intergranular lane.
They suggested that \replaced{the}{a} LOS velocity gradient is one of the major causes of line broadening\added{,
although they did not have a spectral resolution enough to obtain a velocity gradient because of the filtergraph observation}.

\replaced{Our objective is to study line broadening and LOS velocity gradient.}{
There has been little observational knowledge on the contribution made to broaden spectral line profiles by LOS velocity gradients,
which is critical to investigate the small-scale velocity fields.
Our objective is to determine spatial and temporal variations of the line broadening and LOS velocity gradients
with spatial and spectral resolutions enough to resolve the granulation dynamics.
}
These quantities can be effectively derived from spectroscopic observations.
We \replaced{analyze}{analyzed} the data obtained with the Solar Optical Telescope \replaced{(SOT) (\citealt{Tsuneta08})}{(SOT; \citealt{Tsuneta08})}
onboard the {\it Hinode} satellite \citep{Kosugi07};
this instrument is advantageous in providing stable spectroscopic observations with \added{a} high spatial resolution. 
In Section \ref{sec:obana}, the observational data and the analysis methods are described.
The results regarding spatial distributions and temporal evolutions are discussed in Section \ref{sec:results}.
Furthermore, the relationship between line broadening and LOS velocity gradients and small-scale turbulent velocities is discussed in Section \ref{sec:discussion}.

\section{Observation and Analysis}\label{sec:obana}

\subsection{Instruments}
We examined the spectropolarimetric data obtained with the Hinode-SOT.
The SOT has an aperture of 50 cm, thereby attaining a diffraction-limited performance \citep{Suematsu08, Shimizu08}.
The spatial resolution at 6300 {\AA} is approximately $0.^{\prime\prime}3$,
which is equivalent to a distance of about 200 km on the solar surface.
The \replaced{Spectro-polarimeter (SP) (\citealt{Lites+13})}{Spectro-Polarimeter (SP; \citealt{Lites+13})}
covers two neutral iron lines, namely, Fe I 6301.5 {\AA} and 6302.5 {\AA}, with a spectral sampling of 21.5 m{\AA}/pixel.
These two lines arise from the same multiplet,
and their sensitivities to the Zeeman effect are different.
The line-spread function of the SP can be fitted well with a Gaussian shape \citep{Lites+13},
which indicates that the spectral asymmetry caused by the instrumental effects can be neglected.

\begin{table*}[htbp]
  \begin{center}
    \caption{Data sets}
    \begin{tabular}{c|c|c|c|c|c|c|c|c}
      Data Set & Date & $n_x$ & $n_y$ & $n_t$ & $\Delta t$ & $(\delta I)_{\mathrm{rms}}$ & $\delta \Dopfif$ & $\delta \mathrm{FWHM}$\\ \hline
      1 & 22 Nov. 2006 & 1024 & 1024 &  1  & 87 min & 0.47\% &  0.03 km/s & 1.3 m{\AA}\\
      2 & 25 Aug. 2009 &  30  & 384  & 118 & 1 min & 0.96\% &  0.06 km/s & 2.7 m{\AA}\\
      3 & 08 Nov. 2018 &  15  & 512  & 283 & 15 sec & 1.0\% &  0.07 km/s & 2.8 m{\AA}
    \end{tabular}
    \label{tb:data_set}
  \end{center}
\end{table*}

\begin{figure}[t]
\begin{center}
\includegraphics[width=7.5cm]{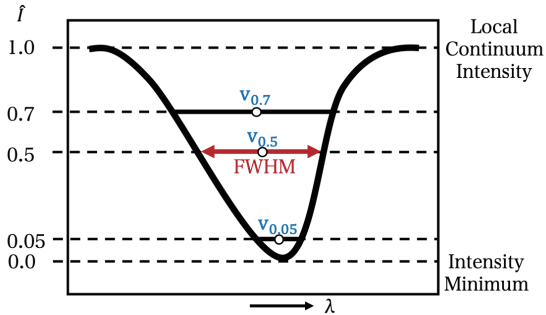}
\end{center}
\caption{Definition of Doppler velocities obtained by bisector analysis.
All the Doppler velocities are defined at specific intensity levels
$\hati=$ 0.05, 0.50, and 0.70, where $\hati$ is the intensity 
normalized by the local continuum intensity and the local minimum intensity of each spectral profile.
FWHM is defined as the full width at the intensity level of $\hati=0.5$.}
\label{fig:def}
\end{figure}

\subsection{Data sets}
Here, two types of data are examined:
normal map,
which includes a single slit-scan with a wide field-of-view (FOV),
and the dynamic mode data,
which repeats slit scans over a narrow FOV
with high time cadence ($\lesssim$ 1 min).
Because the target regions of all the observations are quiet regions near the disk center,
the LOS direction is almost vertical to the surface.
The data sets are represented in Table \ref{tb:data_set}.

The average root-mean-square fluctuation of the continuum intensity $(\delta I)_{\mathrm{rms}}$
provides a photometric error of the data.
The pixel numbers $n_x$ and $n_y$ denote the FOV in \added{the }pixel scales,
where the $x$ and $y$ axes correspond to the scan direction (East-West) and the slit direction (South-North),
with \deleted{a }plate scales of $0^{\prime\prime}.15$ and $0^{\prime\prime}.16$, respectively. 
The number of scans in the dynamic mode is given by $n_t$,
and $\Delta t$ represents the duration of each scan.

These Stokes profiles are calibrated with the standard calibration routine \verb|sp_prep| \citep{Lites_Ichimoto13}.
The \verb|sp_prep| routine includes i) dark-field subtraction, ii) flat-field correction,
iii) calibration of $I, Q, U$, and $V$ crosstalk, iv) correction of curved spectral lines,
v) removal of orbital variation of wavelength and spatial shifts
caused by thermal deformation of instrument optics,
and vi) calibration of intensity variation along the SP slit caused by a small variation in slit width.

The zero-point of the Doppler velocity is adjusted using 
convective blueshift \citep{Dravins81} of
a spectral line profile averaged in each data set.
The Doppler shift of the spatially averaged profile of the Fe I 6301.5 {\AA} line
was accurately determined by \citet{Lohner18}.
They reported that the Doppler shift at the intensity level of
\replaced{$\frac{1}{2}\langle I_{\mathrm{continuum}}\rangle$}{$\frac{1}{2}\langle I_{\mathrm{cont}}\rangle$} is -290 m/s,
where $\langle \cdot \rangle$ denotes the average over the entire FOV\added{ and $I_{\mathrm{cont}}$ is the continuum intensity at each pixel}.
We adjust the zero-point to make our Doppler velocity of the spatially and temporally averaged line profile
at \replaced{$\frac{1}{2}\langle I_{\mathrm{continuum}}\rangle$}{$\frac{1}{2}\langle I_{\mathrm{cont}}\rangle$} be equivalent to -290 m/s.

\subsection{Definition of Parameters}\label{sec:ana_def}
The parameters studied in this paper are summarized in Table \ref{tb:para},
where $\langle \cdot \rangle$ denotes the spatial and temporal average over both the entire FOV and \replaced{all}{the whole} duration in each data set.
All the parameters are calculated for the 6301.5 {\AA} line, except for $\ptot$.
$\icont$ is the continuum intensity normalized by the mean continuum intensity averaged over both the entire FOV and \replaced{all}{the whole} duration,
and $\hati$ represents the intensity normalized by the local continuum and local minimum intensity at each pixel.
\replaced{The}{A} LOS dependence of \replaced{the}{a} Doppler velocity can be estimated by applying bisector analysis (e.g. \citealt{Adam76}) to the spectroscopic data,
by defining three Doppler velocities at $\hati=0.05$, 0.5, and 0.7,
as $\Dopofi$, $\Dopfif$, and $\Dopsev$, respectively (Figure \ref{fig:def}).
Bisector analysis is an appropriate method to extract \deleted{the }LOS velocity \replaced{gradient}{gradients} from \replaced{asymmetricity in a spectral line profile}{asymmetric line profiles}.
The bisector is the midpoint of the same intensity points of both sides of an absorption line.
The Doppler velocity at different intensity levels defined by this bisector method
reflects \replaced{the Doppler}{a LOS} velocity at a different height.
The Doppler velocities at $\hati=0.05$ and $\hati=0.7$ reflect \deleted{the }LOS velocities in the upper photosphere
and the lower photosphere, respectively.
The velocity difference $\veldif$, which is a representative of \replaced{the}{a} Doppler velocity gradient along the LOS,
is defined as $\Dopofi - \Dopsev$.
A positive $\veldif$ indicates strong blueshifts in the lower photosphere,
whereas a negative $\veldif$ indicates strong redshifts in the lower photosphere.
$\delta \Dopfif$, an error of $\Dopfif$, is estimated from $(\delta I)_{\mathrm{rms}}$ (see Appendix {\ref{ap:error}}).
We use the full width at $\hati = 0.5$ (FWHM) as a representative of \added{the }spectral line broadening.
In addition, we use the equivalent width (EW), which is a simple integration of the residual intensity.
Finally, we use the total polarization ($\ptot$)
calculated by the \verb|sp_prep| routine.
$\ptot$ reflects the polarization caused by the Zeeman effect (see Appendix {\ref{ap:ptot}}).

\begin{table}[htbp]
\begin{center}
\caption{Definitions of \added{the }parameters}
\begin{tabular}{|c|c|}\hline
parameter & definition \\ \hline
$\icont$ & $\icont \equiv \frac{I_{\mathrm{cont}}}{\langle I_{\mathrm{cont}}\rangle}$\\
$\hati$ & $\hati \equiv \frac{I-\mathrm{min}(\ilam)}{I_{\mathrm{cont}}-\mathrm{min}(\ilam)}$ (Equation [\ref{eq:I_hat}])\\
$\Dopofi$ & Doppler velocity at $\hati=0.05$\\
$\Dopfif$ & Doppler velocity at $\hati=0.5$\\
$\Dopsev$ & Doppler velocity at $\hati=0.7$\\
$\veldif$ & $\veldif$ $\equiv$ $\Dopofi$ - $\Dopsev$\\
FWHM & Full width at $\hati=0.5$\\
EW & EW $\equiv \int\left( 1-\frac{I_{\lambda}}{I_{\mathrm{cont}}} \right) d\lambda$\\
$\ptot$ & $\ptot \equiv \int \frac{\sqrt{Q_{\lambda}^2+U_{\lambda}^2+V_{\lambda}^2}}{I_{\mathrm{cont}}} d\lambda$\\\hline
\end{tabular}
\label{tb:para}
\end{center}
\end{table}

\begin{figure*}[t]
\begin{center}
\includegraphics[width=18cm]{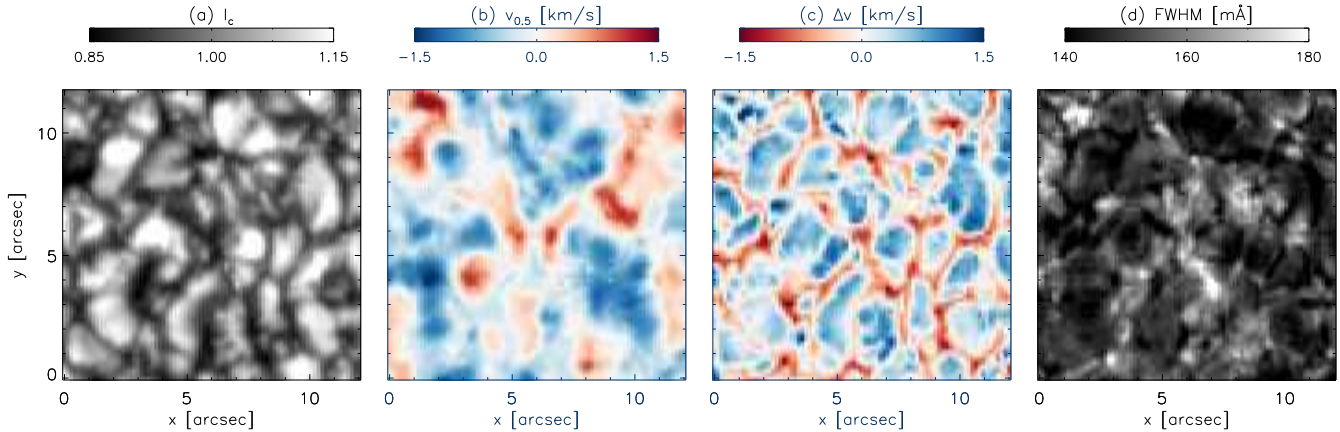}
\end{center}
\caption{Spatial distributions of (a) continuum intensity $\icont$, (b) Doppler velocity $\Dopfif$, (c) velocity difference $\veldif$, 
and (d) FWHM of the Fe I 6301.5 {\AA} line from data set 1.}
\label{fig:map}
\end{figure*}

\section{Results}\label{sec:results}
\subsection{Statistical Analyses of Spectral Line Broadening}\label{sec:hist}
Figure \ref{fig:map} shows the spatial distributions of $\icont$, $\Dopfif$, $\veldif$, and FWHM.
In the spatial distributions of $\icont$ and $\Dopfif$,
cellular patterns related to the granulation can be clearly seen.
The spatial distribution of $\veldif$ shown in Figure \ref{fig:map}(c)
clearly indicates that $\veldif$ is positive within the granules, whereas it becomes negative in the surrounding intergranular lanes.
On the contrary, the FWHM shown in Figure \ref{fig:map}(d) exhibits less clear association with the granular structures.
For example, large granules around $(x,y) = (2^{\prime\prime}, 2^{\prime\prime}.5)$ and $(x,y) = (7^{\prime\prime}.5, 10^{\prime\prime})$
appear to have \deleted{a }small FWHM within the granules, whereas the FWHM \replaced{is}{are} large toward the edges of the granules and in the intergranular lanes.
This result is consistent with the results of previous studies \citep{Gadun97, Hanslmeier08, Khomenko10}.
However, the relationship is not clear for small granules,
e.g., around $(x,y) = (2^{\prime\prime}.5, 7^{\prime\prime})$ and $(x,y) = (8^{\prime\prime}, 7^{\prime\prime})$.
Moreover, sporadic enhancements of FWHM are seen throughout the FOV.

\begin{figure*}[htbp]
\begin{center}
\includegraphics[width=15cm]{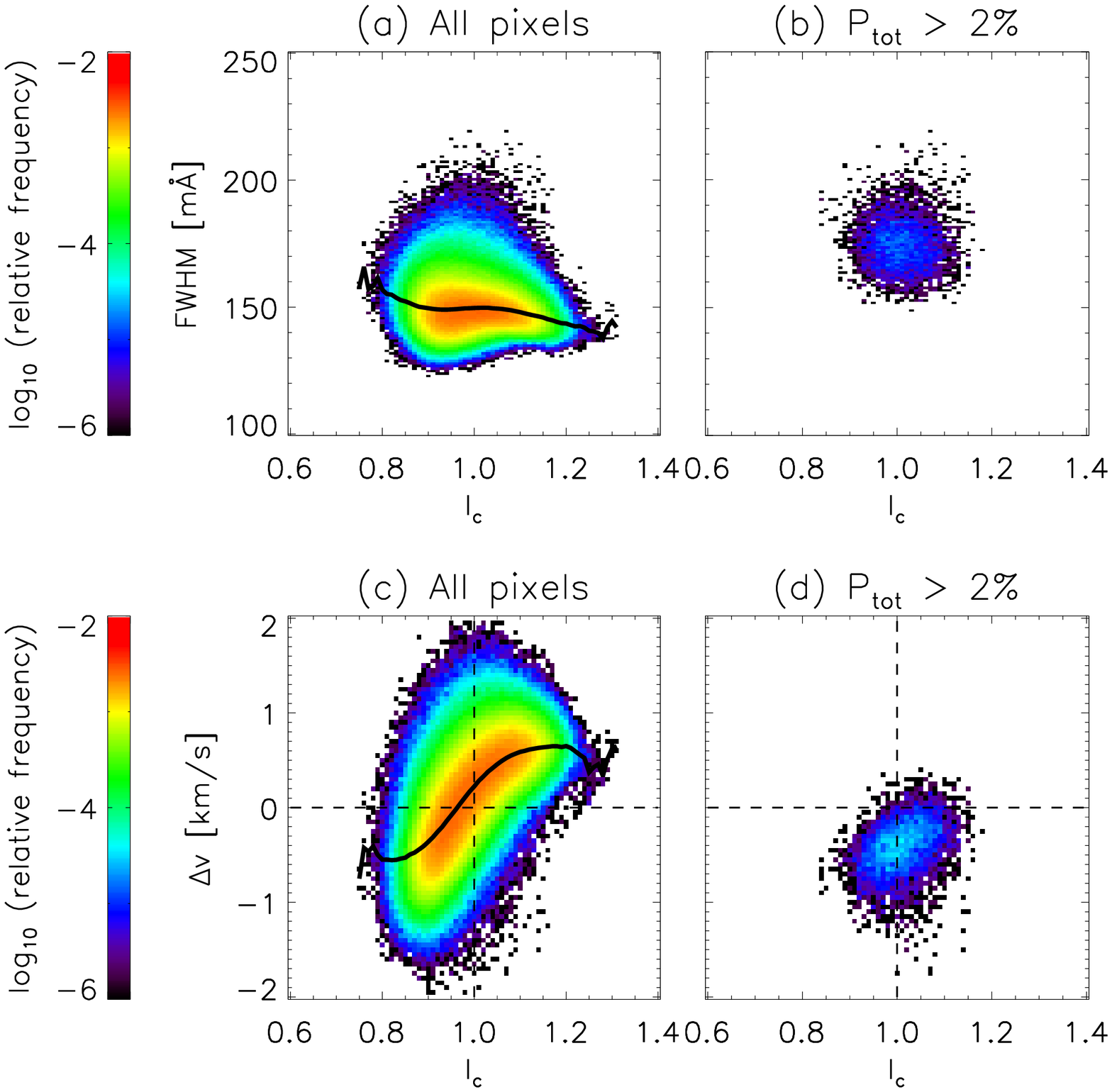}
\caption{2-D histograms of $\icont$ and FWHM \replaced{(panels a and b)}{(panels [a] and [b])},
and $\icont$ and $\veldif$ \replaced{(panels c and d)}{(panels [c] and [d])},
for all pixels in the FOV and the pixels with large \replaced{total polarization}{$\ptot$}.
The $\ptot$ of 2\% corresponds to a magnetic flux of about 450 $\mathrm{Mx/cm^2}$.
The black lines in \replaced{panels a and c}{panels (a) and (c)} indicate the average FWHM and $\veldif$ at each $\icont$.
The color contours show the relative frequency in logarithmic scale.
The histograms are obtained from the data set 1.}
\label{fig:hist1}
\end{center}
\end{figure*}

Figure \ref{fig:hist1}(a) shows a 2-D histogram of
$\icont$ versus FWHM for all pixels of the data set 1.
The black line in Figure \ref{fig:hist1}(a) indicates the average FWHM at each $\icont$.
The trend of the black line indicates that FWHM becomes small in the middle of granules ($\icont > 1.0$),
whereas it becomes large in intergranular lanes ($\icont < 1.0$).
This trend is consistent with the trend reported by previous studies \citep{Gadun97, Hanslmeier08, Khomenko10}.
Furthermore, some pixels with $\icont \sim 1.0$ have \deleted{a }large FWHM and
some with $\icont \sim 0.9$ have \deleted{a }small FWHM.
The deviation of FWHM at each $\icont$ is significantly larger than the estimated error of FWHM caused by the photometric noise of data set 1
(see Table {\ref{tb:data_set}} and Appendix \ref{ap:error}).

Figure \ref{fig:hist1}(c) shows a 2-D histogram of $\icont$ and $\veldif$.
Clearly,\added{ pixels with} negative $\veldif$ exist in the intergranular lanes,
which are caused by \deleted{an }accelerated \replaced{downflow}{downflows}; where the downward velocity becomes faster in the lower layer,
as predicted by the numerical simulations performed by \citet{Gadun97} and  \citet{Asplund00}.
On the contrary,\added{ pixels with} positive $\veldif$ exist near the center of the granules.
These signs of $\veldif$ are consistent with the overshooting convection,
where deceleration of upflows occurs in the upper photosphere.
In addition,\replaced{ a}{ pixels with} strong positive $\veldif$ exist\deleted{s} in the region with $\icont \sim 1.0$.
Such strong positive $\veldif$ has never been indicated by both numerical and observational studies.

\begin{figure*}[htbp]
\begin{center}
\includegraphics[width=15cm]{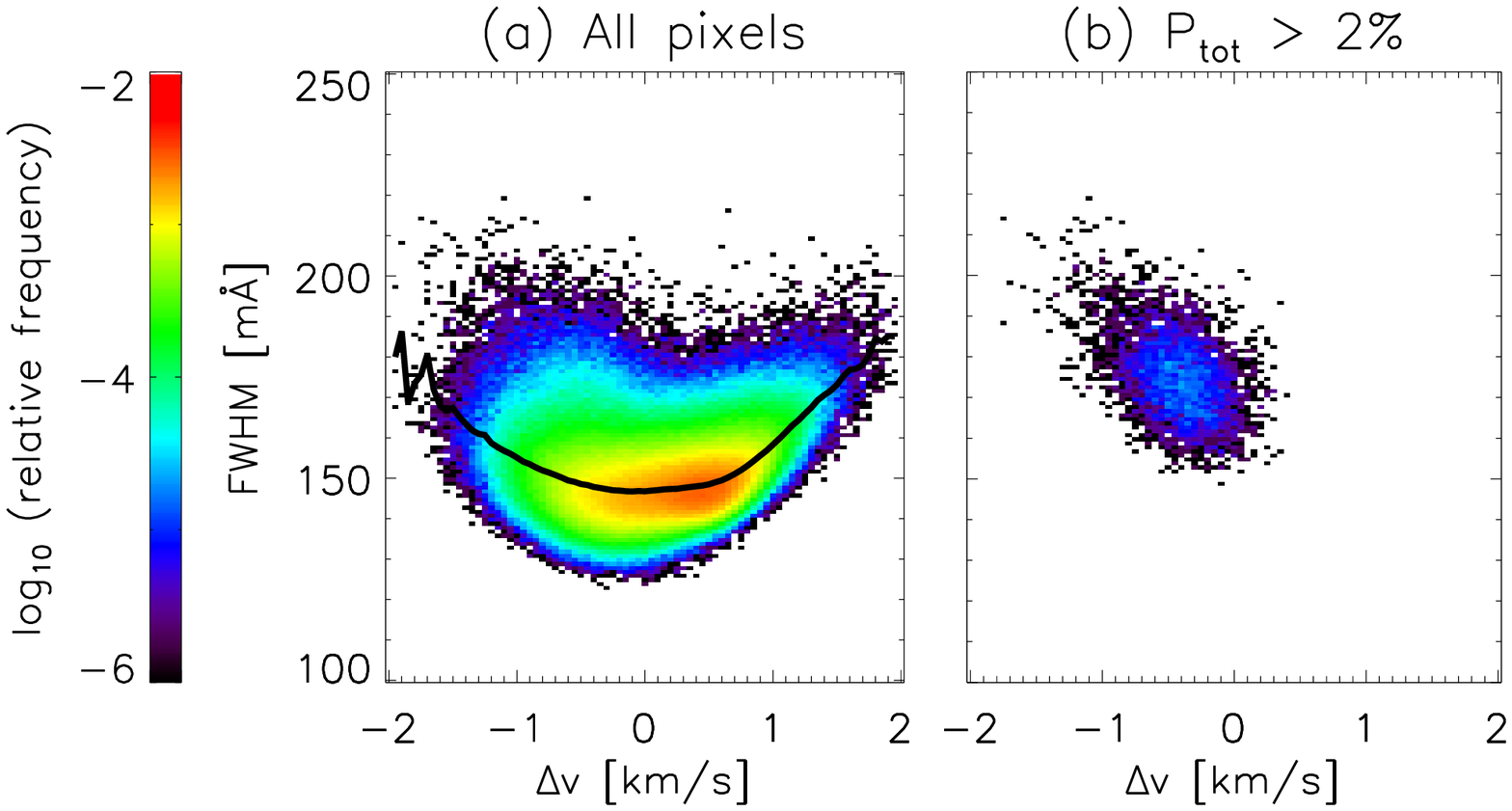}
\caption{2-D histograms of $\veldif$ and FWHM for all the pixels (a) and for those with large total polarization \replaced{(panel b)}{(panel [b])}.
The black line represents the average FWHM at each $\veldif$.
The color contours are the same as those in Figure \ref{fig:hist1}.
The histograms are obtained from the data set 1.}
\label{fig:hist2}
\end{center}
\begin{center}
\includegraphics[width=15cm]{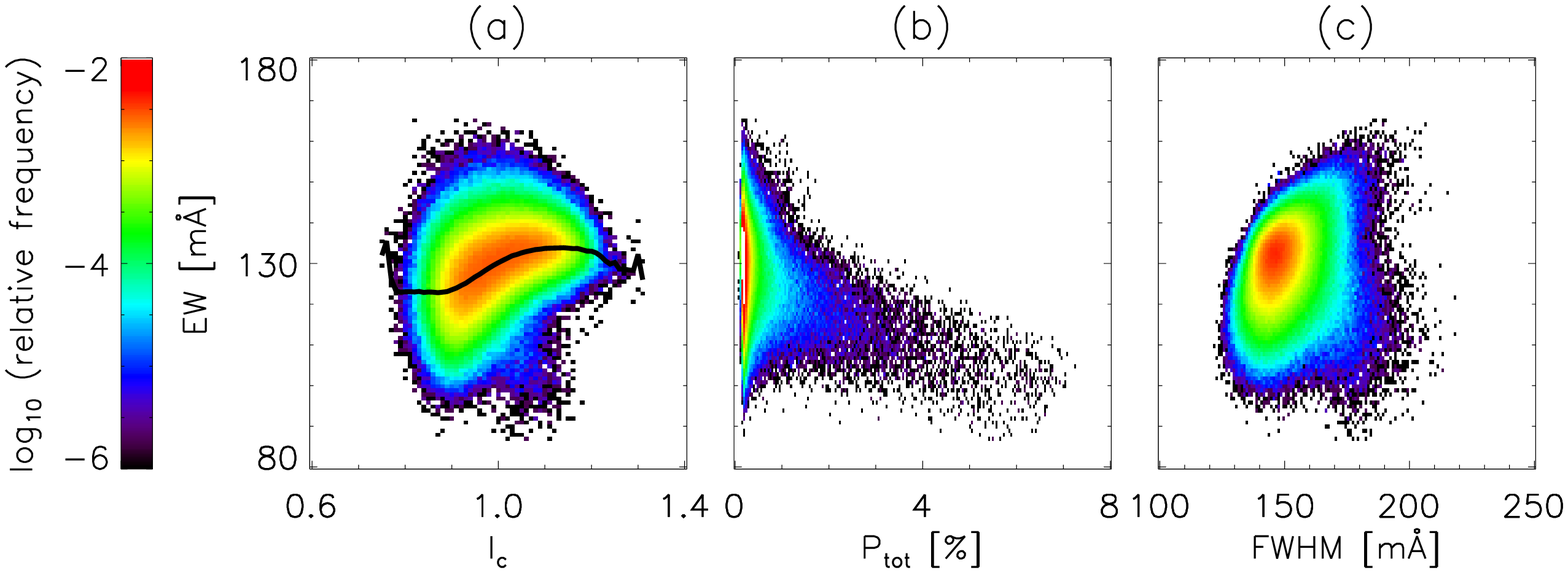}
\caption{(a) Dependence of EW on $\icont$ and (b) \replaced{Total}{total} polarization $\ptot$, 
which is an agent of magnetic flux.
Panel (c) shows the histogram of EW and FWHM.
The color contours are the same as those in Figure \ref{fig:hist1}.
The histograms are obtained from the data set 1.}
\label{fig:hist3}
\end{center}
\end{figure*}

Previous studies have suggested that the spectral line broadening is caused by the LOS gradient of Doppler velocity \citep{Gadun97}.
The relationship between $\veldif$ and FWHM is shown in Figure \ref{fig:hist2}(a).
The figure clearly indicates that large absolute values of $\veldif$ induce an increase of FWHM.
The relationship shown in Figure \ref{fig:hist2}(a) indicates a significant dispersion in FWHM,
implying that another factor also causes line broadening.

It must be noted that the FWHM related to positive $\veldif$ is slightly larger than that of their negative counterparts,
except for the regions where $\ptot$ is large.
The increase in FWHM related to a positive $\veldif$ of 1 km/s is approximately 15 m{\AA} on average,
whereas that related to a negative $\veldif$ of -1 km/s is approximately 10 m{\AA} on average (Figure \ref{fig:hist2}[a]).

We focus on the strongly magnetized regions to understand the effect of magnetic fields.
Figure \ref{fig:hist1}(b) shows the same histogram in panel (a), but for the pixels with strong magnetic flux.
The threshold is defined by the total polarization: $\ptot > 2.0$\%,
which corresponds to magnetic flux \replaced{stronger}{larger} than the equipartition magnetic flux of approximately 450 $\mathrm{Mx/cm^2}$ (see Appendix \ref{ap:ptot}).
Although the magnetic fields are likely to be concentrated in the intergranular lanes,
the continuum intensity in the strongly magnetized region is brighter than that in the weakly magnetized intergranular lane (Figure \ref{fig:hist1}[d]),
because the photospheric surface observed by the continuum intensity moves downward
because of the decrease in opacity caused by magnetic fields \citep{Wilson74, Spruit81}.
Figure \ref{fig:hist2}(b) shows the same histogram in Figure \ref{fig:hist2}(a) but
for the strongly magnetized pixels ($\ptot >$ 2\%).
The figure clearly indicates that the strongly magnetized pixels have large \deleted{values of }FWHM for negative $\veldif$,
but they are not associated with positive $\veldif$.

The histogram of EW and $\icont$ is shown in Figure \ref{fig:hist3}(a).
It can be seen that the EW \replaced{is}{are} small in the intergranular lanes
relative to those in granules.
In addition, in the region of $\icont \sim 1.0$,
the deviation of EW is large.
Figure \ref{fig:hist3}(b) shows the histogram of EW and $\ptot$,
which indicates that the magnetic flux decreases the EW.
We confirmed that the pixels with small EW around $\icont \sim 1.0$
seen in panel (a) are associated with the strongly magnetized pixels.
The strongly magnetized pixels have small EW and large FWHM (Figure \ref{fig:hist3}\added{[b] and }[c]).
The pixels with large EW and $\icont \sim 1.0$ seen in panel (a)
have large FWHM, which is clearly different from the properties of the strongly magnetized pixels.

\begin{figure}[htbp]
\begin{center}
\includegraphics[width=8cm]{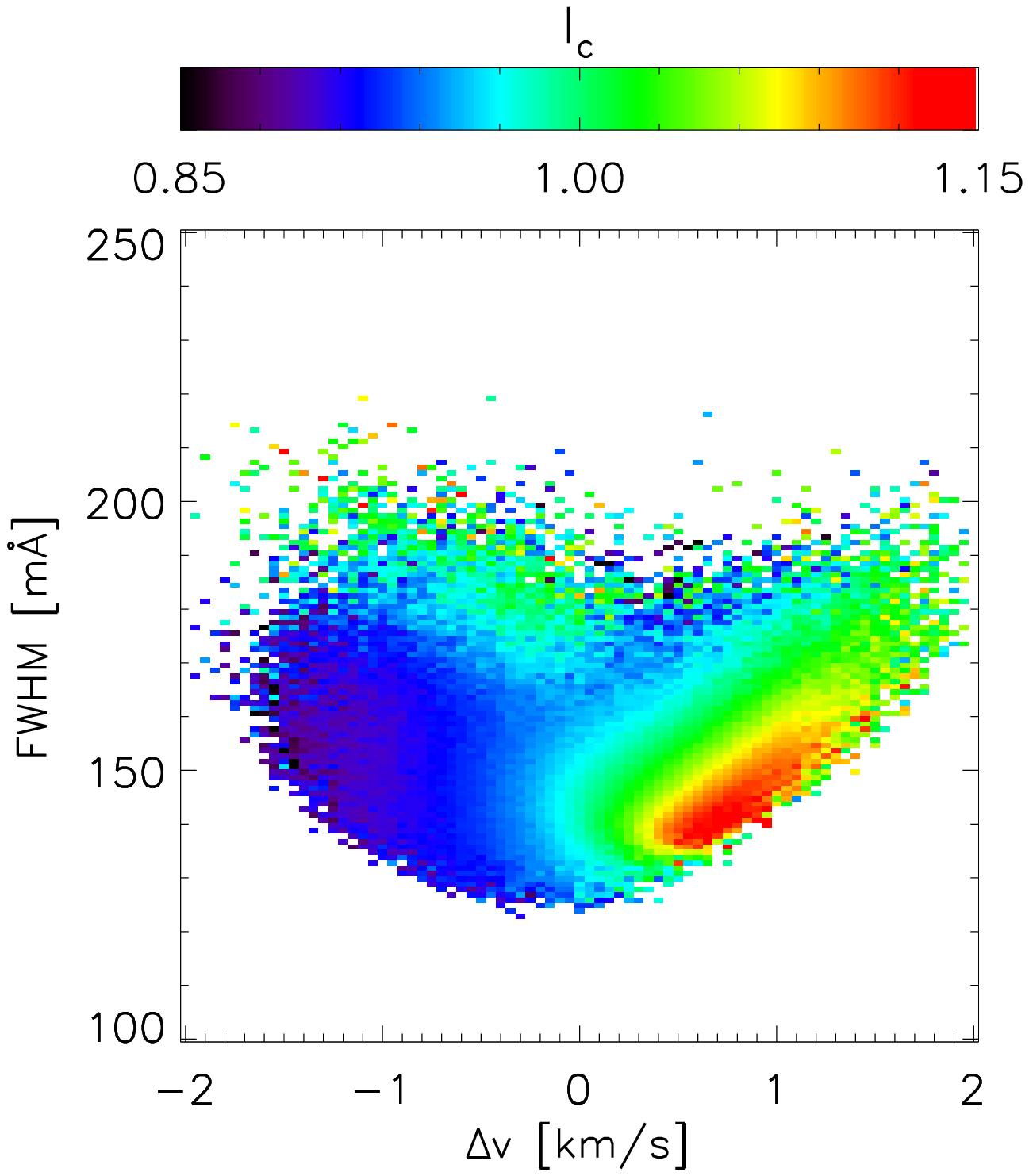}
\caption{Relationship between $\veldif$, FWHM, and $\icont$ from the data set 1.}
\label{fig:fwcorr}
\end{center}
\end{figure}

Figure \ref{fig:fwcorr} shows the relationship between $\veldif$ (abscissa), FWHM (ordinate),
and $\icont$ (color contour),
where the $\icont$ of each bin is obtained by averaging $\icont$ of all the pixels in each bin.
Apparently, near the center of a granule ($\icont \gtrsim 1.1$; highlighted in red),
$\veldif$ \replaced{has a positive value of}{is positive at} approximately 0.8 km/s,
whereas it \replaced{has a negative value of}{is negative at} approximately -1.0 km/s in \replaced{the }downflow \replaced{lanes}{lane} ($\icont \lesssim 0.9$; highlighted in dark blue).
The strongly magnetized pixels are located in the region with $\icont \sim 1.0$ (highlighted in green) and $\veldif < 0$, and have \deleted{a }very large FWHM.
Furthermore, the pixels with $\icont \sim 1.0$ are seen in the region with $\veldif > 0$ as well.
Some of them have \deleted{a }large $\veldif$ and large FWHM, and some have $\veldif \sim 0$ and \deleted{a }small FWHM.
The latter may correspond to the cellular boundary between granules and intergranular lanes, which was found by \citet{Khomenko10}.

\subsection{Case Study}\label{sec:case_study}
\begin{figure*}[tp]
\begin{center}
\includegraphics[width=12cm]{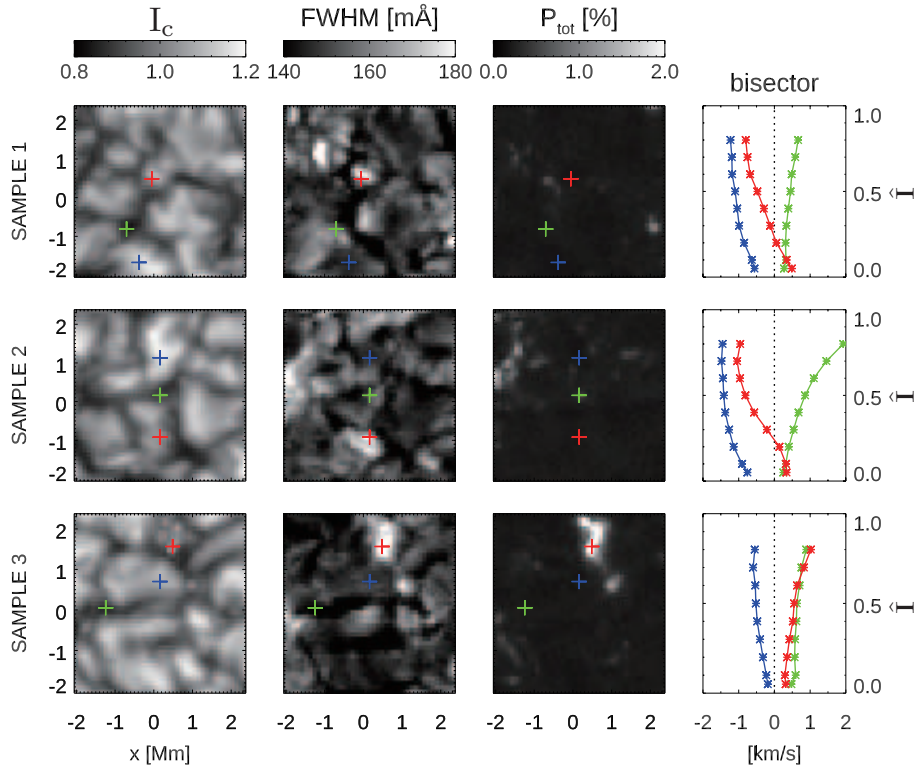}
\caption{Three columns on the left show the spatial distributions of
$\icont$, FWHM, and $\ptot$ from left to right. The rightmost column provides the bisectors
at each pixel indicated by cross symbols in the spatial distributions.
We chose the bisectors in intergranular lanes (green), near the center of granules (blue), and in the region with large FWHM (red).
Note that, in sample 3, the pixel denoted by the red cross symbol is in a
region with strong magnetic flux.
All the samples are extracted from the data set 1.}
\label{fig:event}
\end{center}
\end{figure*}

We examine three different samples in Figure \ref{fig:event},
where the three columns on the left show the spatial distributions of
$\icont$, FWHM, and $\ptot$ (left to right).
The rightmost column provides the bisectors
at each pixel (indicated by cross symbols in the three columns on the left).
The green and blue symbols represent samples of intergranular lanes and center of granules, respectively.
The red symbols represent samples of pixels with large FWHM for sample 1 and sample 2, and show a pixel with strong magnetic flux for sample 3.
The existence of large positive $\veldif$ accompanied by large FWHM without strong magnetic field is seen in samples 1 and 2.
The pixels with large FWHM have $\icont$ of almost unity,
some of which are located at the boundaries between granules and intergranular lanes.
The FWHM of such pixels \replaced{is}{are} larger than \replaced{that}{those} of the pixels in the intergranular lanes or near the center of granules by 30 m{\AA} to 40 m{\AA};
further, even $\ptot$ \replaced{is}{are} small.
The velocity differences $\veldif$ obtained by the \replaced{bisector}{bisectors} (red lines in the rightmost column of Figure \ref{fig:event})
have large positive values compared with those of the other two (green and blue lines in the rightmost column).

Sample 3 shows the case of a strong magnetized region
with large $\ptot$, large FWHM, and large negative $\veldif$.
Although this magnetized region is located in an intergranular lane,
it has a brighter intensity ($\icont \sim 1.0$).
This trend is the same as that shown in panels (b) and (d) in Figure \ref{fig:hist1}.

\added{
In Sections \ref{sec:hist} and \ref{sec:case_study},
we analyzed the spatial distribution of FWHM and newly found the sporadic enhancements of FWHM with positivie $\veldif$ throughout the photosphere.
It was also shown that the large dispersion of FWHM cannot be explained only by the LOS velocity gradient.
To identify the cause of the sporadic line broadening, we analyze the temporal evolution in Section \ref{sec:temporal}.
}

\subsection{Temporal Evolution}\label{sec:temporal}
\begin{figure*}[tp]
\begin{center}
\includegraphics[width=12cm]{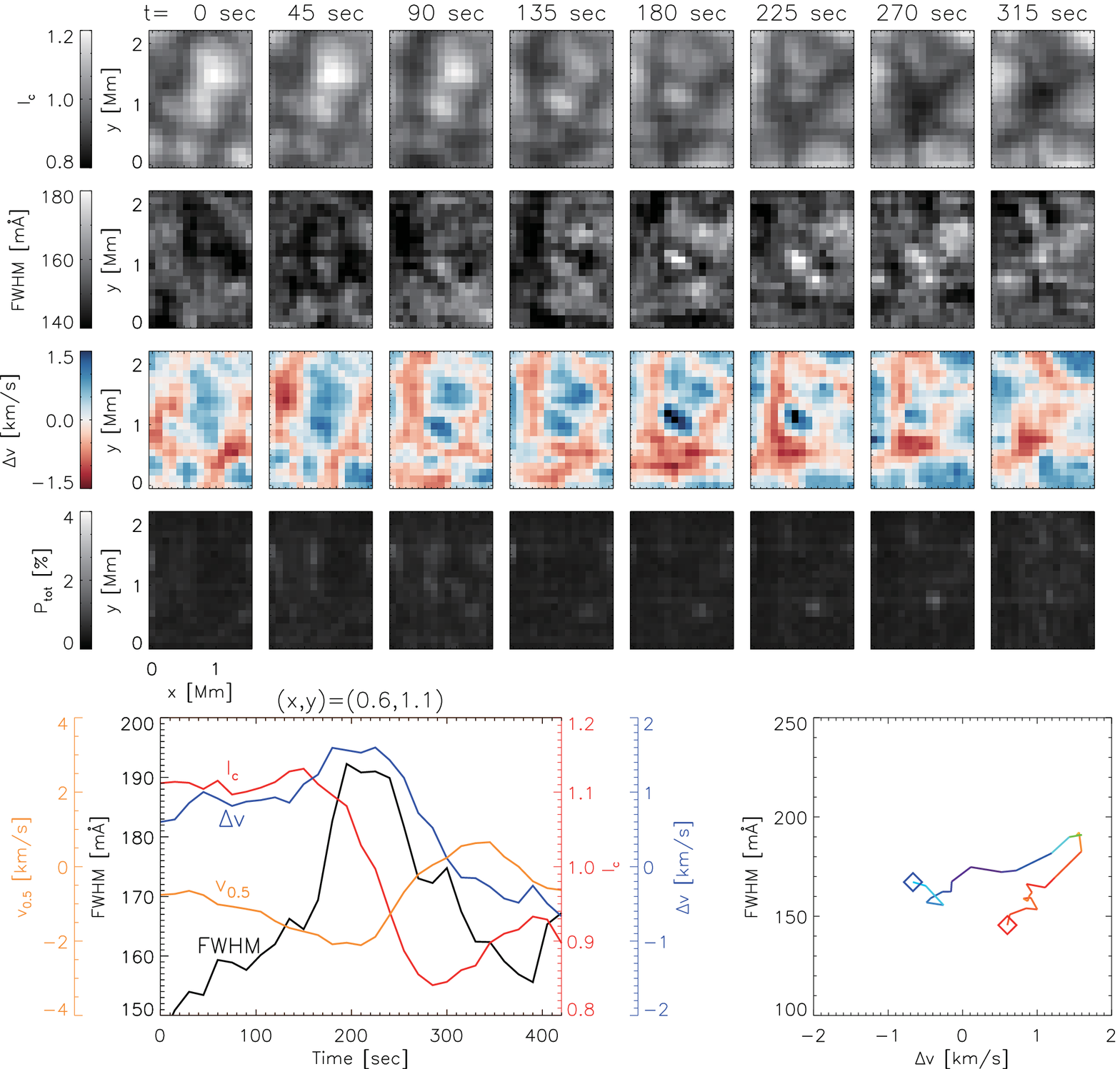}
\caption{A sample of time evolution of a fading granule.
Four rows on the top provide the temporal evolutions of $\icont$, FWHM, $\veldif$, and $\ptot$
(top to bottom), respectively.
The bottom left panel shows the temporal evolutions of FWHM (black), $\Dopfif$ (orange), $\icont$ (red), and $\veldif$ (blue)
of the fading granule at (x,y)=(0.6,1.1).
This temporal evolution is plotted in the $\veldif$-FWHM diagram (right bottom panel)
with color-coding according to the continuum intensity $\icont$ (same as that in Figure \ref{fig:fwcorr}),
where the red and blue diamonds denote the first frame ($t=0$ sec)
and the final frame ($t=420$ sec), respectively.
This sample is extracted from the data set 3.}
\label{fig:temp_ex1}
\end{center}
\end{figure*}

\begin{figure*}[t]
\begin{center}
\includegraphics[width=12cm]{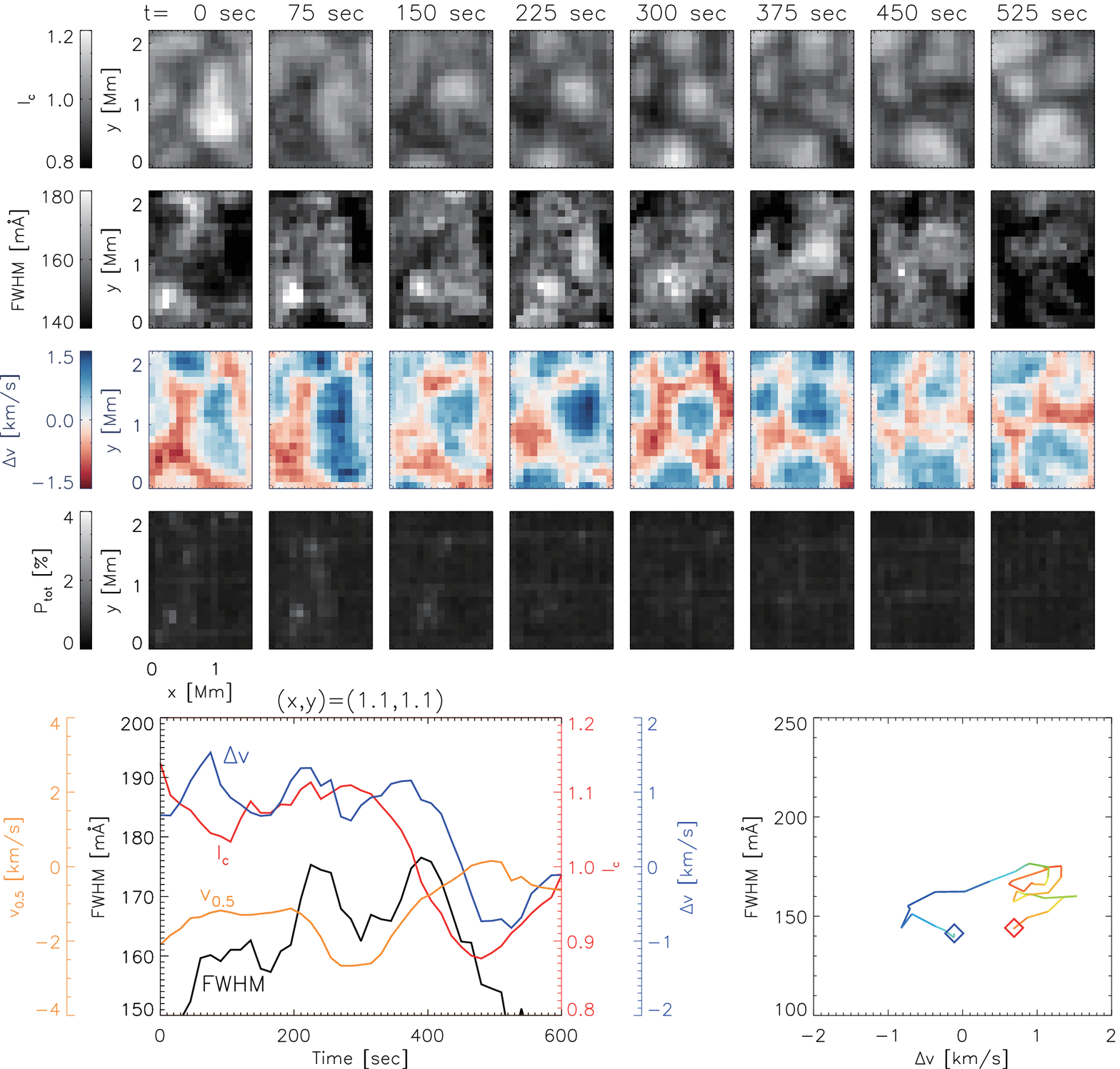}
\caption{Same figure as Figure \ref{fig:temp_ex1} but for another sample.
Note that this sample granule takes \added{a }longer period to fade out than that in Figure \ref{fig:temp_ex1}.
This sample is extracted from the data set 3.}
\label{fig:temp_ex2}
\end{center}
\end{figure*}

We examine the temporal evolution of the granules in the data sets 2 and 3
to investigate development of positive $\veldif$ and
broadening of spectral lines.
Figures \ref{fig:temp_ex1} and \ref{fig:temp_ex2} show two samples of temporal evolution of granules.
In this study, we focus on the fading phase of granules.
The four rows on the top provide the temporal \replaced{evolution}{evolutions} of $\icont$, FWHM, $\veldif$, and $\ptot$
(top to bottom).
The bottom left panel in Figure \ref{fig:temp_ex1} shows the
temporal \replaced{evolution}{evolutions} of FWHM (black), $\Dopfif$ (orange), $\icont$ (red), and $\veldif$ (blue)
at (x,y)=(0.6,1.1),
where the center of the granule is located at t=225 s.
This temporal evolution is plotted in the $\veldif$-FWHM diagram (bottom right panel),
where the red and blue diamonds denote the first (t=0 sec) and the final (t=420 sec) frames, respectively.

In the earlier phase of the sample shown in Figure \ref{fig:temp_ex1} (from 0 s to 150 s),
the continuum intensity does not vary significantly around $\icont\sim1.1$ and the upward flow becomes faster
from -0.5 km/s to -1.5 km/s,
enhancing the positive $\veldif$ and FWHM;
subsequently, $\icont$ starts decreasing.
When $\veldif$ reaches the maximum,
simultaneously, FWHM and blueshift of $\Dopfif$ reach their maxima as well.
Further, $\icont$ starts decreasing from 1.10 to 0.85 within approximately 100 s,
which we define as the fading phase.
Finally, $\veldif$ and FWHM drop, and $\Dopfif$ changes drastically from -1.5 km/s (blueshift) to +1.0 km/s (redshift).
$\ptot$ does not change significantly over this period.

Another sample of a fading granule is given in Figure \ref{fig:temp_ex2}.
The increase in FWHM takes a longer period than that of the sample in Figure \ref{fig:temp_ex1},
and the decrease of $\icont$ takes approximately 150 s.
In the earlier phase, $\icont$ does not change significantly for approximately 400 s.
Although the positive $\veldif$ does not increase, the FWHM increases;
this increase cannot be explained only by the velocity gradient.
Finally, $\icont$, FWHM, and $\veldif$ drop, and
$\Dopfif$ changes its sign;
$\ptot$ does not \replaced{increase}{vary so much} over this period again.

In both the samples, if we plot the temporal evolution in the $\veldif$-FWHM diagram,
it begins from bottom right ($\veldif$$>$0 and small FWHM) and proceeds to top right ($\veldif$$>$0 and large FWHM),
and finally drops to bottom left ($\veldif$$<$0 and small FWHM).
Combining this temporal evolution with Figure \ref{fig:fwcorr}\deleted{(a)},
we can confirm that the regions with $\veldif > 0$ and large FWHM
(top right domain in the $\veldif$-FWHM diagram in Figure \ref{fig:temp_ex1} and \ref{fig:temp_ex2})
are preferentially seen in the fading phase of granules.

\begin{figure}[t]
\begin{center}
\includegraphics[width=9cm]{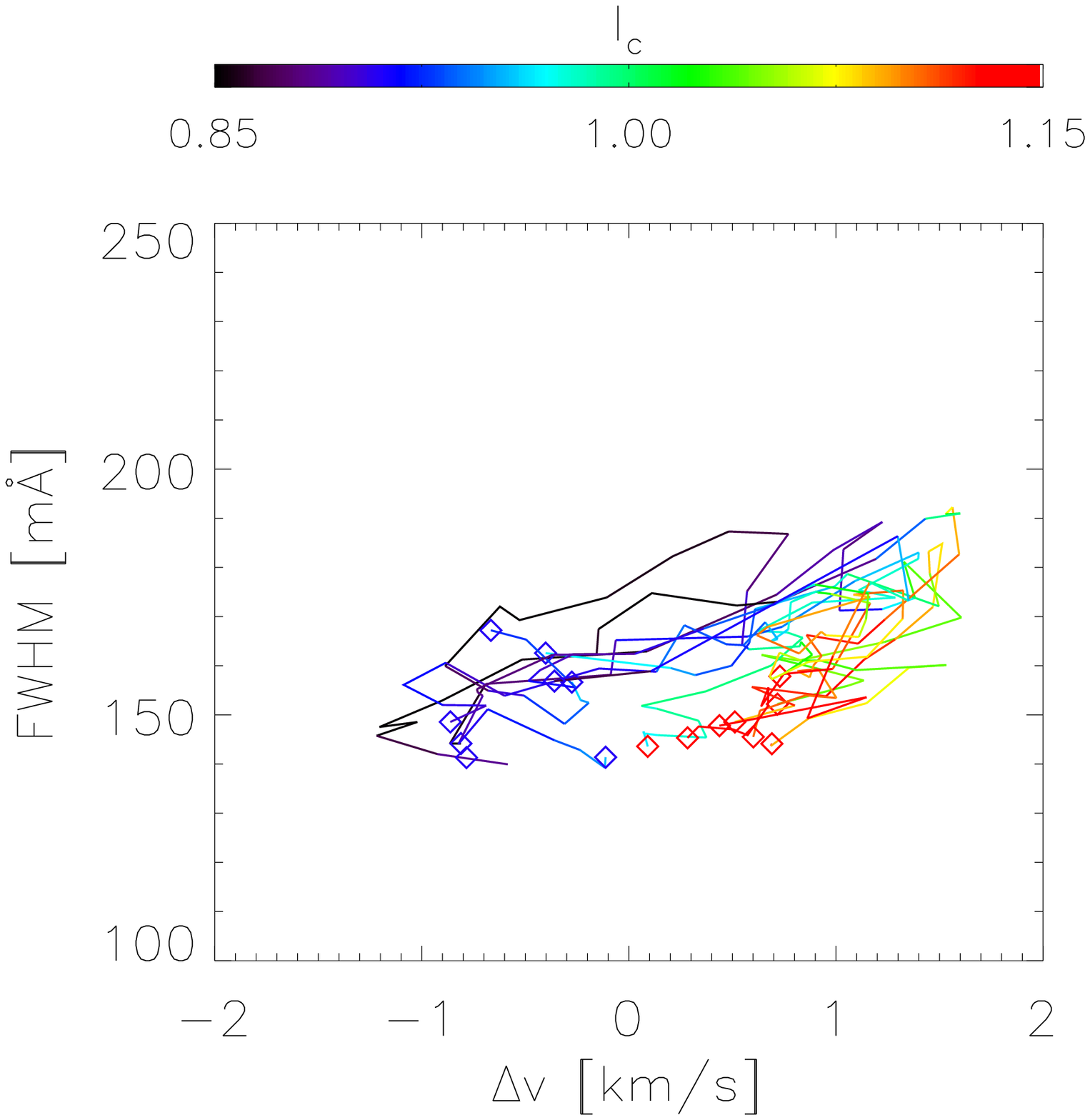}
\caption{Time evolutions of eight sample granules.
Comparison with Figure \ref{fig:fwcorr}\deleted{(a)} is useful.
One sample is from the data set 2 and the other samples are from the data set 3.
The samples \replaced{described}{shown} in Figure \ref{fig:temp_ex1} and \ref{fig:temp_ex2} are included as well.}
\label{fig:temp_sum}
\end{center}
\end{figure}

Figure \ref{fig:temp_sum} shows eight samples of fading granules in the $\veldif$-FWHM diagram.
It is clearly seen that the FWHM and $\veldif$ of all the samples become large,
approximately when $\icont$ becomes almost unity in the fading phase of granules.
As a granule fades out, $\veldif$ decreases and changes its sign.
The FWHM is larger than that in the earlier phase even when $\veldif$ is almost zero.
This demonstrates that the increase in FWHM is caused by another effect rather than the velocity gradient,
which is discussed later in Section \ref{sec:dis_turb}.

\section{Summary \& Discussion}\label{sec:discussion}
\subsection{Summary of the Observation}
We examined both the normal map data and the dynamic mode data obtained with Hinode-SOT
and investigated the nature of spectral line broadening by analyzing the velocity difference $\veldif$
and FWHM of the spectral line profiles.
Further, positive $\veldif$ exists, indicating strong blueshifts in the lower atmospheric layer;
the existence of negative $\veldif$ indicates strong redshifts in the lower layer.
We confirmed that spectral line broadening is induced by the velocity gradient.
As suggested by the numerical simulations,
the downward velocity is higher in the lower layer in the intergranular lanes,
and this velocity gradient increases the FWHM,
whereas the boundaries between granules and intergranular lanes have small FWHM because of small velocity gradients.
We confirmed these characteristics observationally, by analyzing the spectroscopic data with \added{a }high spatial resolution.

We newly found that the regions with large positive $\veldif$ have $\icont$ of almost unity
and very large FWHM, some of which are seen in the fading phase of the granules.
Some regions with negative $\veldif$ are associated with strong magnetic flux.
On the contrary, no clear relationship is observed between positive $\veldif$ and magnetic flux.

\subsection{Spectral Line Broadening caused by Velocity Gradients}\label{sec:dis_vd}
\replaced{The velocity}{Velocity} gradient in \deleted{the }intergranular lanes, where stronger downward flows occur in the lower atmosphere,
\replaced{have already been studied}{was already known in the former studies}.
In this study, we find the significant velocity gradients associated with fading granules,
where gas moves upward in the lower atmosphere and downward in the upper atmosphere.
When a granule fades out, the gas in the upper layer could start descending earlier than that in the lower layer
because the gas in the upper layer is cooled by radiation earlier than that in the lower layer,
giving rise to a large velocity gradient.
The velocity gradients in the intergranular lanes and the fading granules have different signs; however, both increase the spectral line widths (Figure \ref{fig:hist2}\added{[a]}).
The fading granules are mostly located in the less-magnetized regions, and the spectral line broadening has no clear association with the magnetic flux.
This implies that the line broadening seen in the fading granules is not caused by the Zeeman effect.

\subsection{Spectral Line Broadening caused by Turbulent Motions}\label{sec:dis_turb}
Even when $\veldif$ is almost zero, large dispersion is seen in FWHM in Figure \ref{fig:hist2}\added{(a)}.
If we attribute this dispersion of FWHM to small-scale turbulent motions,
the average turbulent velocity is estimated by the following equation:
\begin{equation}
\sqrt{\overline{\mathrm{FWHM}}^2-\left(\overline{\mathrm{FWHM}}-\sigma_{\mathrm{FWHM}}\right)^2}
= 2\sqrt{2\ln{2}} \frac{\overline{v_t}}{c}\lambda \label{eq:vt}
\end{equation}
where $\sigma_{\mathrm{FWHM}}$, $\overline{\mathrm{FWHM}}$, $\lambda$, and $c$ represent
the standard deviation of FWHM at $\veldif=0$ km/s in Figure \ref{fig:hist2}(a),
average FWHM at $\veldif=0$ km/s of 147 m{\AA},
wavelength of 6301.5 {\AA}, and the speed-of-light, respectively.
The average turbulent velocity $\overline{v_{\mathrm{t}}}$ is estimated at 0.9 km/s using the equation (\ref{eq:vt}).

As mentioned in Section \ref{sec:hist}, a difference between the spectral line broadenings exists;
positive $\veldif$ increases the FWHM by a larger degree
than negative $\veldif$.
This difference can be attributed to
the relationship between turbulent motions and velocity gradients.
When $\icont$ becomes almost unity in the fading phase of granules,
the gas moves downward along with a decrease in its volume,
and the surrounding gas converges into the fading granule.
In addition, the positive $\veldif$ in the fading granules
corresponds to upward motions in the lower layer and downward motions in the upper layer,
which may cause a strong shear of vertical flows,
potentially making the gas more turbulent.
As shown in Figure \ref{fig:temp_sum}, when the granules fade out, the FWHM is larger than that in the earlier phase
even if $\veldif$ is almost zero,
which implies that the turbulent motions probably develop in the fading granules.

In this study, we estimated the approximate Doppler velocity gradient by \added{the }bisector \replaced{analyses}{analysis}.
Evaluating the turbulent velocity, considering temperature and velocity gradients,
using a spectral line inversion such as SIR \citep{SIR92} is important to get a more reliable result.
A 4 m solar telescope named Daniel K. Inouye Solar Telescope (DKIST) is expected to begin its observation soon.
The DKIST has an extremely high spatial resolution, which enables us to resolve small-scale complex structures in the photosphere.
The line broadening and the turbulent motions reported in this paper should be investigated with such high-resolution observations.

Hinode is a Japanese mission developed and launched by ISAS/JAXA,
in collaboration with NAOJ as the domestic partner, and NASA and STFC as the international partners.
The support for post-launch operation is provided by JAXA and NAOJ (Japan),
STFC (UK), NASA, ESA, and NSC (Norway).
Further, R.T.I. is supported by JSPS Research Fellowships for Young Scientists.
This work is supported by \replaced{the NINS program for cross-disciplinary study
(Grant Numbers 01321802 and 01311904) on Turbulence, Transport, and Heating Dynamics
in Laboratory and Astrophysical Plasmas: “SoLaBo-X”}{the NINS program for cross-disciplinary study
on Turbulence, Transport, and Heating Dynamics in Laboratory and Solar/Astrophysical Plasmas: ”SoLaBo-X” (Grant Numbers 01321802 and 01311904)}
and by JSPS KAKENHI Grant Numbers JP18H05234 (PI: Y. Katsukawa) and JP19J20294 (PI: R.T. Ishikawa).

\appendix

\section{Error Estimation of Bisector Analysis}\label{ap:error}
\begin{figure}[t]
\begin{center}
\includegraphics[width=12cm]{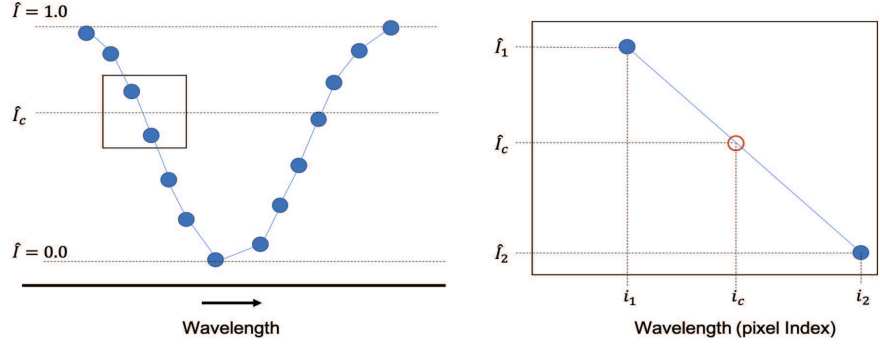}
\caption{Schematic of bisector analysis: blue circles indicate the observed intensity.
To derive the spectral line profile at a specific intensity $\hat{I}_{\mathrm{c}}$,
we interpolate the observational data with the linear functions.
The errors transported into the final results originate from the intensity fluctuation
at both sides of the fitting region, $(i_1,\hat{I}_1)$ and $(i_2,\hat{I}_2)$,
where $i_2=i_1+1$ in \replaced{a}{the} pixel unit.}
\label{fig:bisec_err}
\end{center}
\end{figure}

Here, we describe the error estimation of the bisector analysis.
We define the normalized intensity $\hat{I}$ as,
\begin{equation}
\hati \equiv \frac{I-\mathrm{min}(\ilam)}{I_{\mathrm{cont}}-\mathrm{min}(\ilam)}. \label{eq:I_hat}
\end{equation}
\replaced{Further, we}{We} define $i_c$ as the target spectral position in the pixel unit as follows:
\begin{equation}
i_c \equiv i_2 - \Delta i.
\end{equation}
By interpolating with a linear function (Figure \ref{fig:bisec_err}), we obtain,
\begin{eqnarray}
\Delta i : (i_2-i_1) &=& (\hati_c-\hati_2):(\hati_1-\hati_2), \\
\therefore \Delta i &=& \frac{\hati_c-\hati_2}{\hati_1-\hati_2}.
\end{eqnarray}
If we assume that the error is chiefly caused by the random intensity fluctuation
($|\delta I|/I \equiv \alpha$), the average fluctuation of $\hati$ can be approximately written as follows:
\begin{eqnarray}
|\delta \hati| &=& \frac{|\delta I|}{I_{\mathrm{cont}}-\min(\ilam)}\\
&=& \frac{\alpha I}{I_{\mathrm{cont}}-\min(\ilam)}\\
&=& \alpha\left( \hati + \frac{\min(\ilam)}{I_{\mathrm{cont}}-\min(\ilam)}\right)
\sim (\hati+0.46)\alpha,
\end{eqnarray}
Then, we obtain,
\begin{eqnarray}
\delta(\Delta i) &=& \sqrt{\left(\frac{\partial(\Delta i)}{\partial \hati_2}\right)^2 |\delta \hati_2|^2
+\left(\frac{\partial(\Delta i)}{\partial \hati_1}\right)^2 |\delta \hati_1|^2}\\
&<& \frac{1}{|\hati_1-\hati_2|}\sqrt{|\delta \hati_1|^2+|\delta \hati_2|^2}
= \frac{\sqrt{2}}{|\hati_1-\hati_2|}|\delta \hati|.
\end{eqnarray}
The difference $|\hati_1-\hati_2|$ varies based on $\hati$:
$|\hati_1-\hati_2|\sim0.15$ for $\hati=0.5$,
and $|\hati_1-\hati_2|\sim0.06$ for $\hati=0.05$.

The Doppler velocity is calculated as,
\begin{equation}
v_{\hati} = \frac{\left\{\frac{1}{2}(i_{\mathrm{c,left}}+i_{\mathrm{c,right}})-i_0\right\}\Delta\lambda}{\lambda_0} \ c,
\end{equation}
and the full width is given by,
\begin{equation}
\mathrm{FW}_{\hati} = (i_{\mathrm{c,right}} - i_{\mathrm{c,left}}) \Delta \lambda,
\end{equation}
where $i_0$ and $\Delta\lambda$ represent the pixel index of the line center of Fe I 6301.5 {\AA}
and the spectral sampling (21.5 m{\AA}/pixel) of Hinode-SOT SP, respectively.

For the data set 1, which has an intensity fluctuation of $\alpha=$ 0.47\%,
the errors in Doppler velocities and FWHM are defined as follows:
\begin{eqnarray}
\delta \left(\Dopfif\right)
&= \left.\frac{\delta(\Delta i)}{\sqrt{2}} \left(\frac{\Delta\lambda}{\lambda_0}\right)c \right|_{\hati=0.50}= 0.03 \ \mathrm{km/s}\\
\delta \left(\FWfif\right)
&= \left.\sqrt{2}\delta (\Delta i) \Delta\lambda \right|_{\hati=0.50}= 1.3 \ \mathrm{m{\AA}}\\
\delta \left(\Dopofi\right)
&= \left.\frac{\delta(\Delta i)}{\sqrt{2}} \left(\frac{\Delta\lambda}{\lambda_0}\right)c \right|_{\hati=0.05}= 0.04 \ \mathrm{km/s}
\end{eqnarray}

\section{ Behavior of Total Polarization}\label{ap:ptot}
\begin{figure}[h]
\begin{center}
\includegraphics[width=10cm]{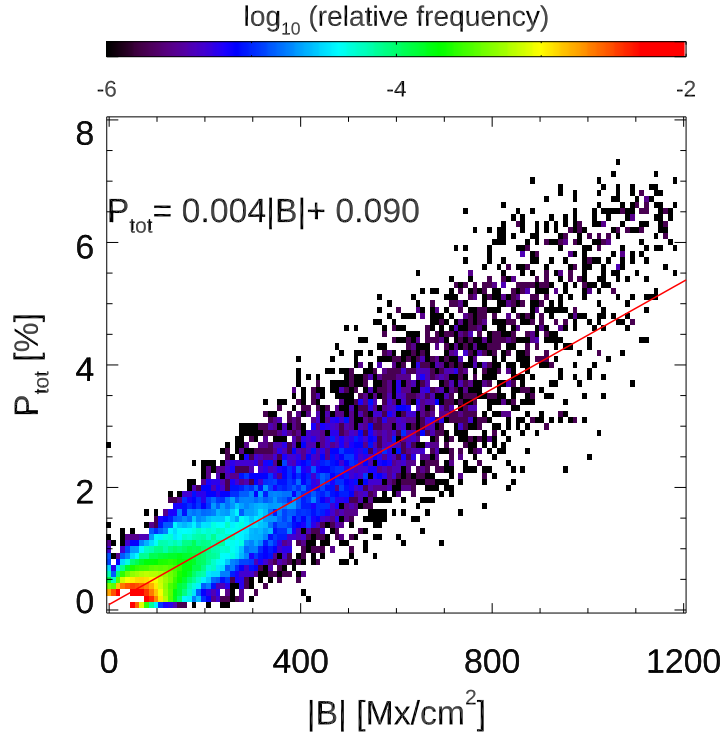}
\caption{Relationship between total polarization $\ptot$ and magnetic flux of the Hinode level 2 data.
The result of linear regression is indicated by the red line.}
\label{fig:ptot_mag}
\end{center}
\end{figure}

We use the total polarization $\ptot$ derived by the \verb|sp_prep| routine \replaced{\citet{Lites_Ichimoto13}}{\citep{Lites_Ichimoto13}}
as the agent of unsigned magnetic flux $|B|$.
\replaced{Here, we describe the relationship between $\ptot$ and $|B|$ (Figure \ref{fig:ptot_mag}).
Further}{In Figure  \ref{fig:ptot_mag}}, we compare the $\ptot$ with $|B|$ of the Hinode level 2 data,
which is derived using the Milne-Eddington approximation.
Then, we obtain a linear relation as follows:
\begin{equation}
\ptot \ [\%] = 0.004 |B| + 0.09.
\end{equation}
\replaced{Further, the}{The} equipartition magnetic flux of approximately 450 $\mathrm{Mx/cm^2}$ corresponds to $\ptot = 2$\%.

\end{document}